\documentclass{ws-mpla}

\usepackage{xspace}
\usepackage{rotating}
\usepackage{psfrag}

\begin{document}

\newcommand{\umin}{\ensuremath{\,u_{\rm min}}\xspace}
\newcommand{\Dl}{\ensuremath{D_{\rm l}}\xspace}
\newcommand{\Dls}{\ensuremath{D_{\rm ls}}\xspace}
\newcommand{\Ds}{\ensuremath{D_{\rm s}}\xspace}
\newcommand{\tE}{\ensuremath{t_{\rm E}}\xspace}
\newcommand{\tN}{\ensuremath{t_{\rm N}}\xspace}
\newcommand{\rhoStar}{\ensuremath{\rho}\xspace}
\newcommand{\Ml}{\ensuremath{M_{\rm l}}\xspace}
\newcommand{\Mp}{\ensuremath{M_{\rm p}}\xspace}
\newcommand{\tzero}{\ensuremath{t_{0}}\xspace}
\newcommand{\rE}{\ensuremath{R_{\rm E}}\xspace}
\newcommand{\thetaE}{\ensuremath{\theta_{\rm E}}\xspace}
\newcommand{\Amax}{\ensuremath{A_{\rm max}}\xspace}

\newcommand{\Mj}{\ensuremath{M_{\rm J}}\xspace}
\newcommand{\Mearth}{\ensuremath{M_{\oplus}}\xspace}
\newcommand{\Msun}{\ensuremath{M_{\odot}}\xspace}
\newcommand{\Mstar}{\ensuremath{M_{\star}}\xspace}

\newcommand{\Rsun}{\ensuremath{R_{\odot}}\xspace}
\newcommand{\Rs}{\ensuremath{R_{\rm s}}\xspace}

\newcommand{\moaLIII}{OGLE-2003-BLG-235/MOA-2003-BLG-53\xspace}
\newcommand{\ogleLXXI}{OGLE-2005-BLG-071\xspace}
\newcommand{\ogleCCCXC}{OGLE-2005-BLG-390\xspace}
\newcommand{\ogleCLXIX}{OGLE-2005-BLG-169\xspace}
\newcommand{\machoXCVIII}{MACHO-98-BLG-35\xspace}

\newcommand{\chisq}{\ensuremath{\chi^{2}}\xspace}

\newcommand{\mas}{{\rm mas}\xspace}
\newcommand{\muas}{\ensuremath{\mu\rm as}\xspace}
\newcommand{\kpc}{\ensuremath{\rm kpc}\xspace}
\newcommand{\mps}{\ensuremath{\rm m\,s^{-1}}\xspace}
\newcommand{\kmps}{\ensuremath{\rm km\,s^{-1}}\xspace}

\newcommand{\etal}{\emph{et~al.}\xspace}

\newcommand{\tabStrut}{\rule[-8pt]{0pt}{20pt}}
\newcommand{\arrStrut}{\rule[-8pt]{0pt}{20pt}}

\input{aas_macros.sty}

\markboth{Rattenbury}
{Planetary Microlensing: From Prediction to Discovery}

\catchline{}{}{}{}{}

\title{ \MakeUppercase{Planetary Microlensing:}\\\MakeUppercase{From Prediction to Discovery}}

\author{\footnotesize NICHOLAS JAMES RATTENBURY}

\address{Jodrell Bank Observatory, Department of Physics \& Astronomy,\\ The University of Manchester, Macclesfield, Cheshire SK11 9DL, \\ United Kingdom \\njr@jb.man.ac.uk}

\maketitle

\pub{Received (Day Month Year)}{Revised (Day Month Year)}

\begin{abstract}
Four planets have recently been discovered by gravitational microlensing. The most recent of these discoveries is the lowest-mass planet known to exist around a normal star. The detection of planets in gravitational microlensing events was predicted over a decade ago. Microlensing is now a mature field of astrophysical research and the recent planet detections herald a new chapter in the hunt for low mass extra-solar planets. This paper reviews the basic theory of planetary microlensing, describes the experiments currently in operation for the detection and observation of microlensing events and compares the characteristics of the planetary systems found to date by microlensing. Some proposed schemes for improving the detection rate of planets via microlensing are also discussed.

\end{abstract}

\ccode{PACS nos.: 97.82.-j, 95.85.Kr, 95.55.-n, 07.87.+v}

\section{Introduction}
The discovery that extra-solar planets are common in the Galaxy ranks as one of the most significant astrophysical discoveries from the last 15 years. Three successful methods for detecting extra-solar planets; pulsar timing\cite{1992Natur.355..145W}, radial velocity measurements\cite{2004A&A...415..391M,2005ApJ...619..570M} and transit observations\cite{2002AcA....52....1U,2003Natur.421..507K,2004A&A...421L..13B,2004ApJ...613L.153A} have yielded diverse planetary systems. To these three successful methods we must now add a fourth: microlensing.

\section{Gravitational Microlensing}	

\begin{figure}[th]
\psfrag{Observer}[][]{Observer}
\psfrag{Lens plane}[][]{\rule{10pt}{0pt}Lens plane}
\psfrag{Source plane}[][]{\rule{10pt}{0pt}Source plane}
\psfrag{Source star}[][]{Source star}

\psfrag{xlabel}[][]{\rE \rule{0pt}{20pt}}
\psfrag{ylabel}[][]{\rE}
\psfrag{umin}[][]{\umin \rule[-20pt]{0pt}{20pt}}
\psfrag{beta}[][]{$\beta$}
\psfrag{d}[][]{$d$}
\centerline{
\begin{tabular}{lr}
\psfig{file=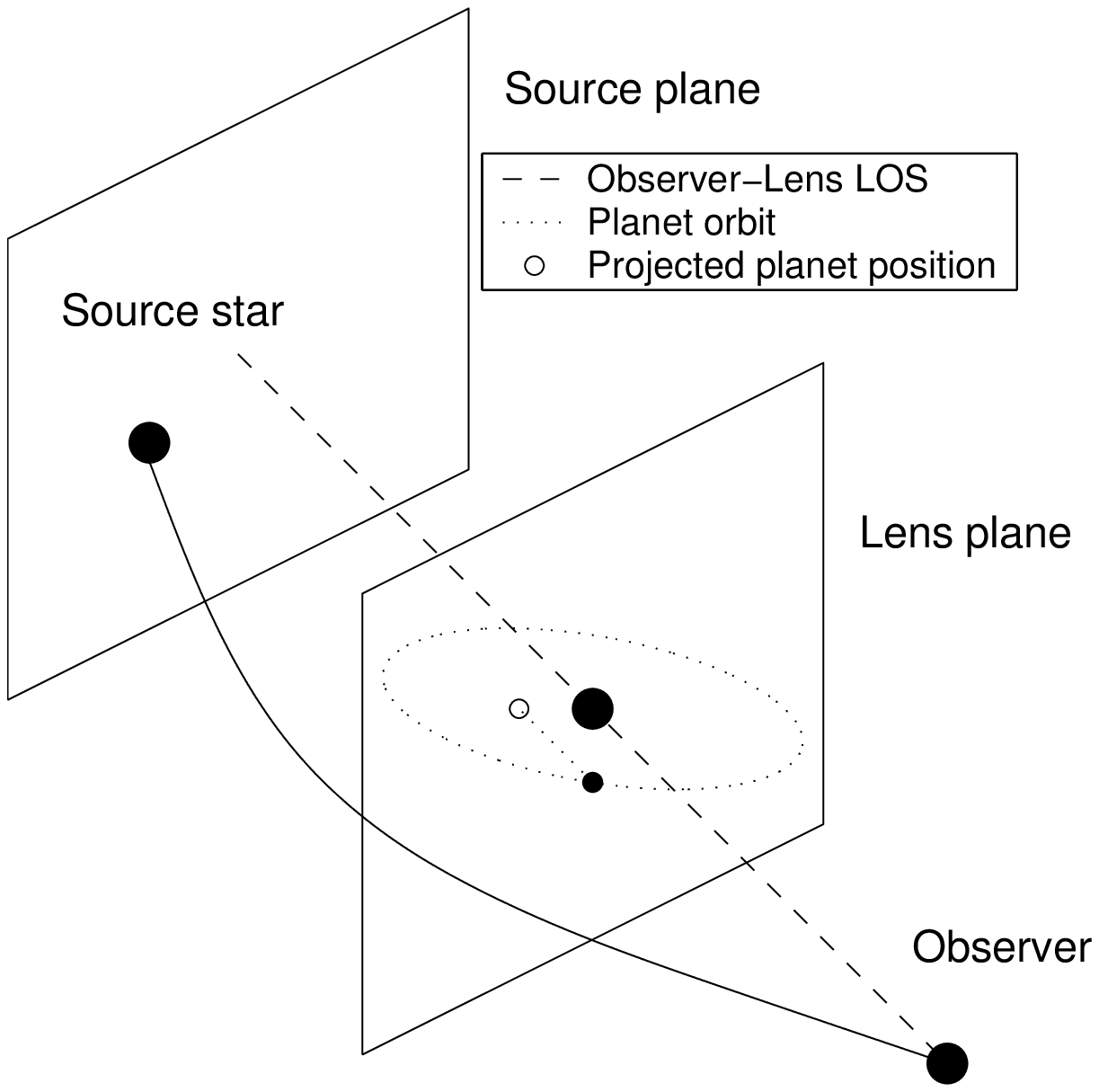,width=0.5\textwidth} &
\psfig{file=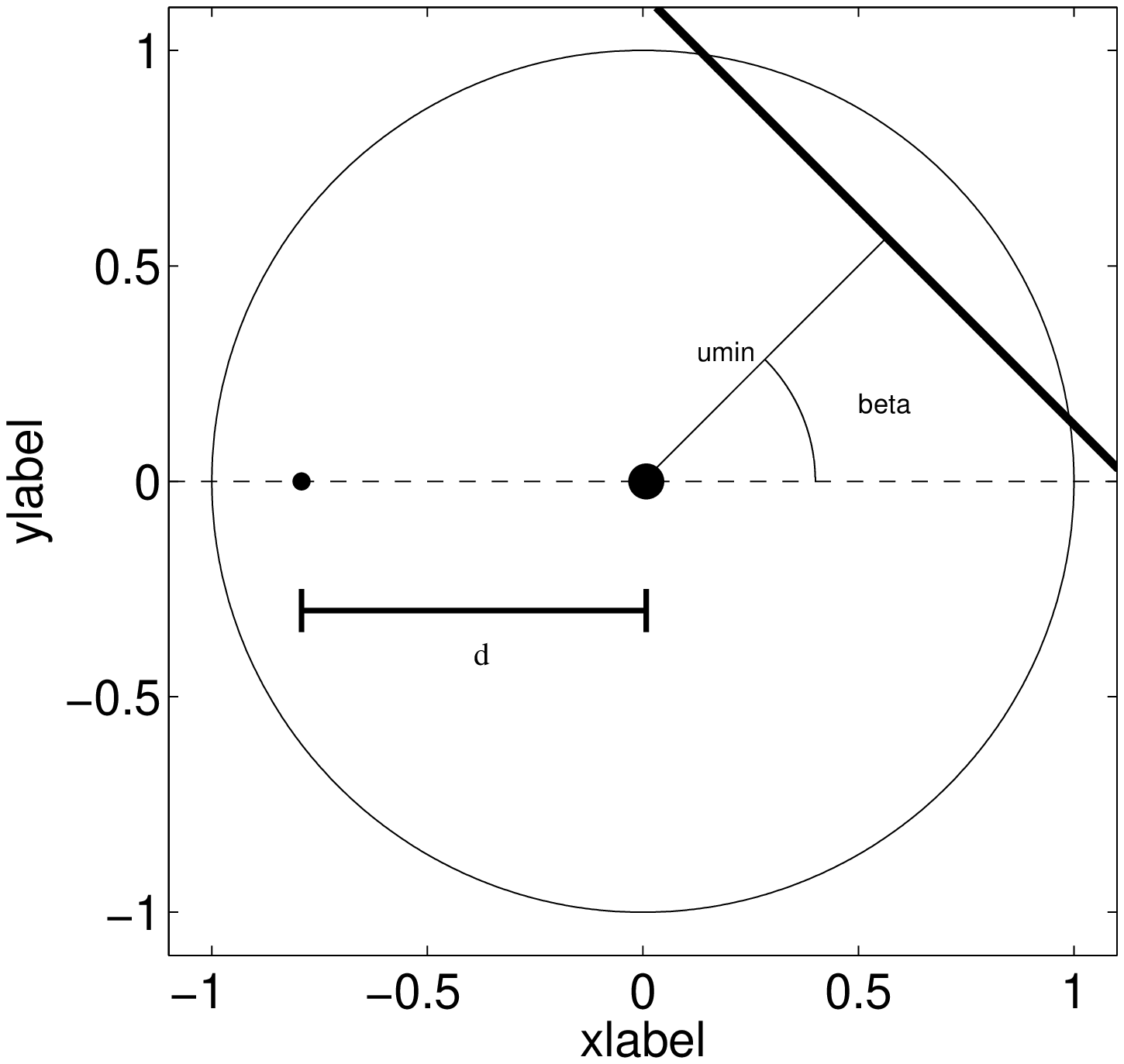,width=0.5\textwidth} \\
\end{tabular}}
\vspace*{8pt}
\caption{Microlensing geometry. Left: Light from a background source star is deflected by the gravitational potential of the lens system, which may comprise a star and planet. Right: The co-ordinate system used here is centred on the centre-of-mass of the lens system, and is in units of the Einstein radius, \rE. The filled black circles denote the lens star and planet. The projected star-planet separation is $d$.  The source moves along a path (heavy line), with minimum approach distance to the CoM of \umin. The binary lens axis is aligned with the horizontal axis, and the line normal to the track of the source star is oriented at angle $\beta$. \protect\label{fig:planes}}
\end{figure}

The phenomenon of gravitational microlensing occurs when light from a background star (the source) is deflected by the gravitational potential of an intervening massive object (the lens)\cite{1964PhRv..133..835L}, see Fig.~\ref{fig:planes}. The lensing action produces at least two images of the source star. For lensing events within our Galaxy, these images cannot be resolved by current telescopes; hence the term \emph{micro}lensing, as distinct from the lensing of background quasars by foreground galaxies which produces resolvable images. The Very Large Telescope Interferometer, however, is capable of resolving the individual microlens images\cite{2006MNRAS.365..792R,2001A&A...375..701D,2003ApJ...589..199D}.  The precision astrometry from the future Space Interferometry Mission (SIM) and \emph{Gaia} space telescopes will be sufficient for detecting the shift in image centroid during a microlensing event\cite{2000ApJ...534..213D}.

In the case of Galactic microlensing, the source and lens objects are typically stars and are in motion relative to observers on Earth. While separate images of the background source star cannot be resolved, the lensing effect produces a time-dependent amplification of the source star. The approach and passage of the lens star close to the source-observer line of sight produces a well-known symmetric light-curve\cite{1986ApJ...304....1P}. 

Planets orbiting the lens star can produce a brief perturbation away from the single-lens light curve\cite{1991ApJ...374L..37M,1992ApJ...396..104G,1994ApJ...436..112B,1997MNRAS.284..172W,2004ASPC..321...47W}. This is very sensitive to the mass of the planet relative to the parent star. Microlensing is sensitive to lens system planets with masses as low as that of the Earth\cite{1996ApJ...472..660B}.

\subsection{Microlensing light curves}
The natural unit of length in microlensing is the Einstein radius:
\begin{equation}
\rE = \left(\frac{4G\Ml}{c^2} \frac{\Dl (\Ds-\Dl)}{\Ds}\right)^{1/2} = 4.42\, {\rm AU} \left(\frac{\Ml}{0.3 M_\odot}\right)^{1/2} \left(\frac{\Ds}{8\kpc}\right)^{1/2} \left( x(1-x) \right)^{1/2},
\label{eq:rE}
\end{equation}
where \Ml is the lens mass, $\Dl$ and $\Ds$ are the distances to the
lens and source respectively, and $x=\Dl/\Ds$. 

Since the lens object, observer and source star are all in relative motion, the projected impact parameter changes with time. This  produces a time-dependent amplification of the source star. The co-ordinate system for a microlensing event is typically placed in the lens plane, with the source star moving behind the lens, see Fig.~\ref{fig:planes}. The amplification of the source star  any time $t$ produced by a single object acting as a gravitational lens is:
\begin{equation}
A(t) = \frac{u^2 + 2}{u \sqrt{u^2 +4}} \hspace{2cm} u(t) = \left[\umin^2 + \left(\frac{v_{\perp} (t - \tzero)}{\rE}\right)^2\right]^{\frac{1}{2}}
\end{equation}
where $u(t)$ is the time-dependent impact parameter, $v_{\perp}$ is the lens transverse velocity with respect to the observer-lens line of sight, $\umin$ is the minimum impact parameter in units of the Einstein radius and \tzero\ is the time of maximum amplification.

The Einstein ring crossing time $\tE = \rE/v_{\perp}$ is the natural time unit for microlensing and can be shown to be:
\begin{equation}
\tE = 38.25 \, {\rm days} \left(\frac{\Ml}{0.3M_{\odot}}\right)^{1/2} \left(\frac{\Ds}{8\, {\rm kpc}}\right)^{1/2}\left((1-x)x\right)^{1/2} ,
\end{equation}
if the transverse velocity of the lens is taken to be $v_{\perp} = 200$ \kmps.

\subsection{Planetary Microlensing Light curves}
The range of signals caused by lens system planets is large, showing a remarkable diversity in shape and duration. The nature of the planetary signal depends on the ratio of the planet mass to the host star, $q=\Mp/\Ml$ and $d$, the orbit radius projected onto the lens plane, in units of \rE (see Fig.~\ref{fig:planes}). The geometry of the microlensing event in terms of the minimum impact parameter, \umin and relative orientation of the source star motion relative to the star-planet axis similarly has a strong effect on the light curve. A planetary microlensing event falls into one of two broadly defined groups, based on when and how the planet perturbation is produced. 

There are positions in the source plane for a binary lens event where the background source star is strongly amplified by the gravitational potential of the lens system. These positions take the form of closed curves called caustics. Whenever the source star crosses a caustic line, a pair of images appears or vanishes depending on whether the source is moving into or out of a region bounded by caustic curves. There is always a small central caustic located near the lens-observer line-of-sight. One or two planetary caustics are also present. The size, orientation and position of the caustics depend on the star-planet mass ratio and projected orbit radius. When a source star passes across a caustic line, sharp changes in source amplification are seen. These \emph{caustic-crossing} events comprise the first type of microlensing events, and the light curves have relatively large and distinctive features. The strong amplification of the source star in a caustic crossing event offers a powerful probe into the nature of both the lens and source star systems\cite{2005A&A...439..645R}.

\begin{figure}[th]
\centerline{
\psfig{file=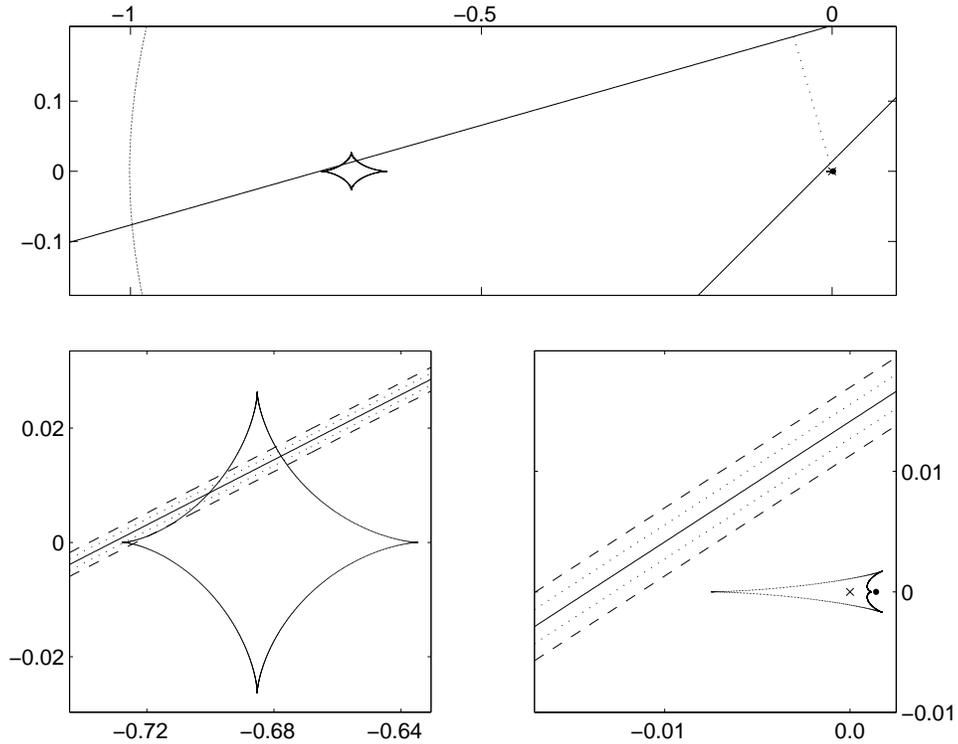,width=1.0\textwidth}
}
\vspace*{8pt}
\caption{Comparison between detecting planets through caustic crossing events and high amplification events. Both axes are in units of \rE. Top: Critical and caustic curves for a planetary lens system with $q = 10^{-3}$ and $d=1.4$. Two source star tracks are shown, one corresponding to a low amplification planetary caustic crossing event, with $\umin = 0.2$, $\beta = 106^{\circ}$; the other corresponding to a high amplification event, passing close to the small central caustic  $\umin = 0.01$, $\beta = 135^{\circ}$. Bottom: Close-up of the planetary caustic (left) and the central caustic (right), with the corresponding source star tracks. Two values of source star radius are used, $\rhoStar = 2\times10^{-3}$ (dashed lines) and $\rhoStar = 1\times10^{-3}$ (dotted lines). The corresponding light curves are shown in Fig.~\ref{fig:himagVsCaustic}.\protect\label{fig:twomodes}}
\end{figure}

\begin{figure}[th]
\psfrag{xlabel}[][]{\tN}
\psfrag{ylabel}[][]{Amplitude}

\centerline{
\psfig{file=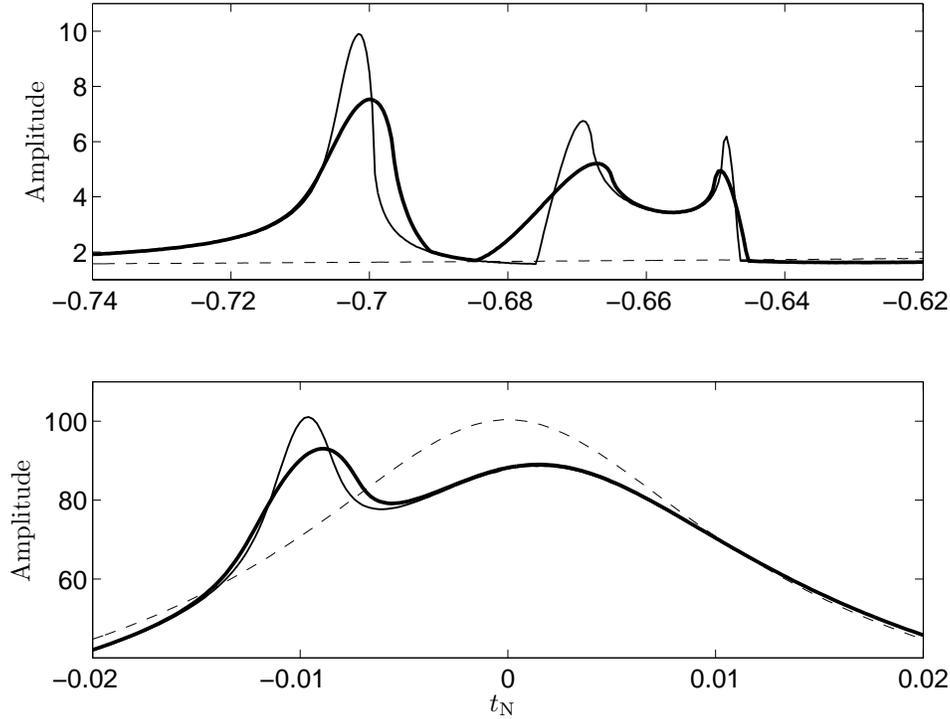,width=1.0\textwidth}
}
\vspace*{8pt}
\caption{Comparison between planetary microlensing light curves in the caustic crossing and high amplification events described in the caption to Fig.~\ref{fig:twomodes}. The vertical axes indicate source amplification and the horizontal axes show normalised time $\tN = (t_{\rm JD} - \tzero) / \tE$. The light curves corresponding to  $\rhoStar = 2\times10^{-3}$  and $\rhoStar = 1\times10^{-3}$ are shown as thick and thin lines respectively. Top: Perturbations corresponding to a caustic cusp crossing ($\tN \simeq -0.7$) and a caustic crossing ($-0.68 \lesssim \tN \lesssim -0.63$). Bottom: Planetary perturbation in the high amplification event.\protect\label{fig:himagVsCaustic}}
\end{figure}

The second class of planetary microlensing events involves the source star passing close to the central caustic. Approaching the cusp of two caustic lines also produces a perturbation in the source amplification, which can be used to infer the presence of the planet\cite{1998ApJ...500...37G}. The perturbations produced in this manner are smaller than those in caustic crossing events, yet because the source passes close to the central caustic, the minimum impact parameter \umin is small and the source is therefore highly amplified. Such \emph{high-amplification} events form the second class of events by which lens system planets can be detected. Fig.~\ref{fig:twomodes} shows an example of the central and planetary caustics for a given microlensing event. Two source star tracks are shown in Fig.~\ref{fig:himagVsCaustic}, one corresponding to a low amplification, planetary caustic crossing event; the other corresponding to a high amplification event with a planetary perturbation near the time of maximum source amplification. Fig.~\ref{fig:himagVsCaustic} shows the caustic crossing and high amplification light curves respectively for these two different types of planetary microlensing event.

Extra-solar planets have been detected in both types of microlensing events. For a detailed comparison between both methods, see Ref.~\refcite{2002MNRAS.335..159R} and Ref.~\refcite{2002MNRAS.331L..19B}.

\subsection{Multiple lens systems}

The effect of a lens system comprised of $n$ objects with masses $m_{i}$ is described via the \emph{lens equation} (following Ref.~\refcite{1999A&A...348..311B}):
\begin{equation}
\label{eq:lenseq}
\mathbf{y} = \mathbf{x} - \sum^{n}_{i=1} m_{i}\frac{\mathbf{x} - \mathbf{x}_{i}}{|\mathbf{x} - \mathbf{x}_{i}|^{2}}
\end{equation}
where $\mathbf{x} = (x_{1},x_{2})$ and $\mathbf{y} = (y_{1},y_{2})$ denote positions in the lens and source planes respectively (see Fig.~\ref{fig:planes}) and the $i$-th mass is located at position $\mathbf{x}_{i}$. Given a source position, $\mathbf{y}$, the values of $\mathbf{x}$ which satisfy Eq.~\ref{eq:lenseq} are the image locations, $\mathbf{x}_{j}$. The \emph{critical curves} are defined as the set of points in the lens plane for which the Jacobian determinant of the lens mapping vanishes. The \emph{caustic curves} are the set of points in the source plane to which the critical curves are mapped via Eq.~\ref{eq:lenseq}:
\begin{center}
\begin{tabular}{cc}
\parbox{150pt}{
$\mathbf{x}_{\rm critical} = \{\mathbf{x} : \det \mathbf{J} = 0\}$ \\
$\mathbf{y}_{\rm caustic} = \{\mathbf{y} : f(\mathbf{x}_{\rm critical})\}$
}
&
\parbox{100pt}{
\[
\mathbf{J}(\mathbf{x})  = \left( \begin{array}{cc}\frac{\partial y_{1}}{\partial x_{1}}\arrStrut & \frac{\partial y_{1}}{\partial x_{2}}\arrStrut \\\frac{\partial y_{2}}{\partial x_{1}}\arrStrut & \frac{\partial y_{2}}{\partial x_{2}}\arrStrut
 \end{array}
\right)
\]
}
\end{tabular}
\end{center}
where $f(\mathbf{x})$ is the lens equation, Eq.~\ref{eq:lenseq}. The amplification of the source star by the lens system is given by $A = \sum A_{j}$ where  $A_{j}$ is  the amplification of the $j$-th image and $A_{j}  = |\det \mathbf{J}(\mathbf{x}_{j})|^{-1}$. From these expressions, it is seen that the caustic curves are the loci of all positions of the source star for which the amplification becomes formally infinite. Such blinding infinities do not occur in Nature, as the ``formally infinite'' amplification just described does not account for the finite extent of the source star\cite{1994ApJ...430..505W}, and does not apply the wave nature of optics.

\section{Survey and Follow-up Collaborations}

The probability that a background star in the Galactic bulge is being microlensed at any time is very small: $\sim 10^{-6}$. In order to observe significant numbers of microlensing events in a given time, millions of stars must be constantly monitored. Two groups currently make routine observations of crowded stellar fields toward the Galactic bulge and Magellanic Clouds. The OGLE collaboration uses a 1.3 m telescope at La Silla, Chile\cite{2003AcA....53..291U}. The MOA collaboration has recently added a 1.8 m telescope as its main survey instrument at Mount John, New Zealand\cite{MOA2pre}. The original 0.6 m survey telescope\cite{2001MNRAS.327..868B} is now used as a follow-up instrument. Both the MOA and OGLE collaborations analyse images in real-time and list on-going events on the internet\footnote{OGLE: http://www.astrouw.edu.pl/$\sim$ogle/ogle3/ews/ews.html\\\phantom{$^{a}$}MOA: http://www.massey.ac.nz/$\sim$iabond/alert/alert.html}. Attention is drawn to events which show clear deviations from a single lens light curve profile. 

Several collaborations perform follow-up observations of interesting microlensing events. The PLANET\cite{1998ApJ...509..687A}, $\mu$FUN\cite{2004ApJ...616.1204Y} and RoboNET\cite{2005DPS....37.3101B} collaborations have telescopes spaced longitudinally around the world, and perform follow-up observations of microlensing events detected by the survey collaborations MOA and OGLE. 

\section{\moaLIII}

Microlensing event OGLE-2003-BLG-235 was identified by the OGLE collaboration on June 22, 2003.  The MOA collaboration independently identified the same event on July 21, 2003 and designated it MOA-2003-BLG-53. The discovery of a planetary signal in this event was presented in Ref.~\refcite{2004ApJ...606L.155B}.

\subsection{Observations}

\begin{figure}[th]
\centerline{
\psfig{file=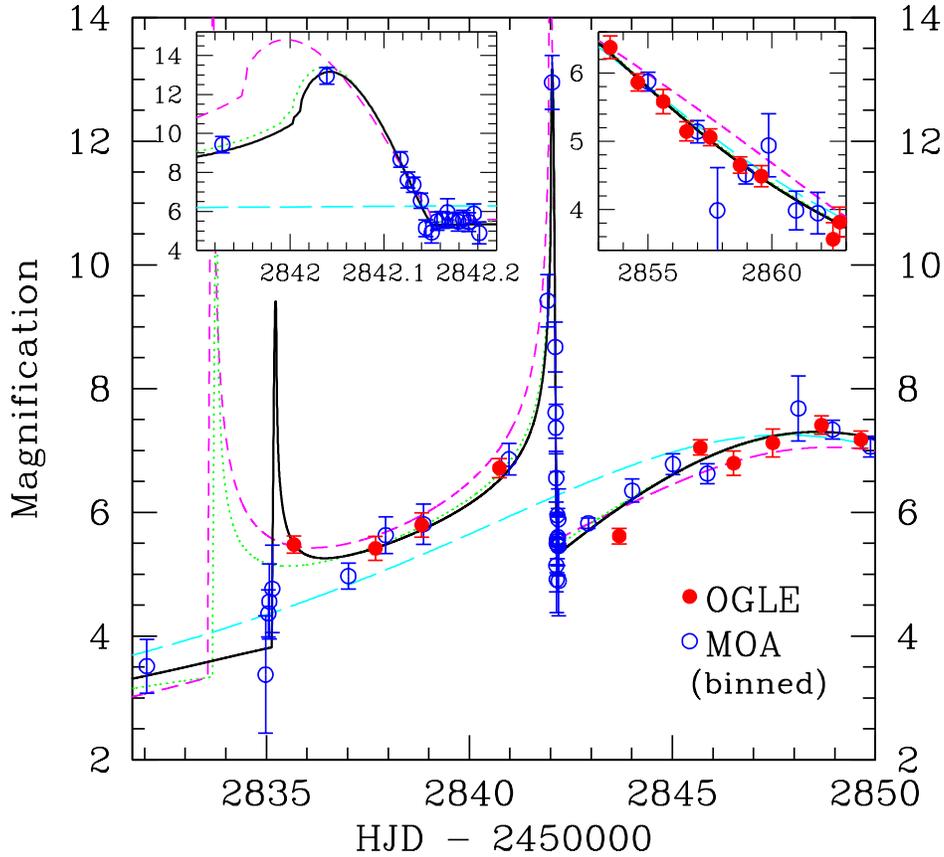,width=1.0\textwidth}}
\vspace*{8pt}
\caption{Planetary microlensing event \moaLIII. The complete dataset is comprised of 183 OGLE and 1092 MOA $I$ band measurements over the 2000-2003 period. A clear caustic crossing signature is seen in the data. Several classes models were used to fit the data; an early caustic crossing model (green/dotted), a non-planetary model, i.e. $q>0.03$ (magenta/dashed) and the best-fitting model (black/solid). The best-fitting single lens model is also shown (cyan/long dash). Reprinted from Ref.~\protect\refcite{2004ApJ...606L.155B}, courtesy D. Bennett.\label{fig:moa53}
}
\end{figure}

Both survey teams observed a seven-day caustic crossing feature which was clear evidence of a binary lens. The MOA collaboration obtained data that closely traced one of the caustic ``spikes''. The nature of the caustic crossing indicated an extreme mass ratio binary lens system. Fig.~\ref{fig:moa53} shows the photometry and best-fitting planetary model for this event during the caustic crossing.



\subsection{Light curve modelling}

Theoretical models of the light curve were produced using three different methods. These methods all converged to the best-fitting model shown in Fig.~\ref{fig:moa53}. The fit parameters used were \umin, \tE, \tzero, the star-planet mass ratio $q$, the projected binary separation, $d$, the source star size (as a fraction of \rE), $\rhoStar = \Rs/\rE$ and the position angle of the binary axis with respect to the source star track, $\phi$. In addition to these seven physical parameters, there were two linear scaling parameters for each passband. 

The model fitting algorithms sought minima in \chisq by varying all parameters simultaneously. The parameter space was thoroughly searched and several different physical models were considered, particularly those which had larger (non-planetary) mass ratios.  Another class of models was found which had an early first caustic crossing. Table~\ref{tab:params} lists the parameters of the best-fitting models for the various classes of microlensing models.

\begin{table}
\scriptsize
\tbl{\label{tab:params}Best fitting model parameters for event \moaLIII}
{\begin{tabular}{ccccccccc} \toprule \toprule
  Model &
  $\Mp/\Ml$ &
  $\rho$ &
  $d$ &
  $\phi$ &
  $\umin$ &
  $\tzero\,^{\rm a}$ &
  $\tE$ &
   $\chisq$ \\
   &
   $\times 10^{-3}$&
    $\times 10^{-4}$&
   &
    (deg)&
   &
   &
  (days) & 
  (1267 dof)\\ \toprule

Best Fit$^{\rm b}$ & $3.9^{+1.1}_{-0.7}$ & $9.6(1.1)$ &
        $1.120(7)$ & $223(1.4)$ & $0.133(3)$ &
        $48.06(13)$ & $61.5(1.8)$ & $1390.49$  \\
Early Caustic & $7.0$ & $10.4$ & $1.121$ & $218.9$ & $0.140$ &
        $47.90$ & $58.5$ & $1397.87$ \\
Best Non-planet & $30.0$ & $8.8$ & $1.090$ & $187.9$ & $0.144$ &
        $46.20$ & $57.5$ & $1601.44$  \\
Single Lens & -- & -- & -- & -- & $0.222$ &
        $47.77$ & $45.2$ & $2041.45$ \\ \botrule
\end{tabular}}
\vphantom{40} $^{\rm a}$Times for \tzero are HJD - 2452800.0. $^{\rm b}1\sigma$ errors.
\end{table}

\subsection{Parameter constraints}

For most microlensing events, the Einstein ring radius crossing time, \tE,  is the only measurable microlensing parameter. The lens mass, lens distance and transverse velocity are three-fold degenerate, and usually cannot be resolved uniquely for a given event. Statistical arguments based on galactic models can give some estimates of these parameters, yet do not constrain the degenerate parameters strongly. It is  possible however, to determine the finite size of the source star in some binary lens caustic crossing events. Light curve modelling gives $\rho$, the source star size as a fraction of the angular Einstein radius. The source angular size can be inferred using the source magnitude and color and bulge population data. This leads to an estimate of the event Einstein ring radius, and thereby, the degeneracy between \Ml, \Dl and $v_{\perp}$ can be partly resolved. 

The  color and baseline magnitude of the source star indicated a G-type bulge star near the MS turn-off. Using color-color relations, and empirical relations between $V-K$ and surface brightness, the source angular radius, $\theta_{\rm s}$, was determined.  The measured value of $\rho = \theta_{s}/\thetaE$  combined with the estimate of $\theta_{\rm s}$ gave $\thetaE = 520 \pm 80$ \muas. 
Using this value for \thetaE  and Eq.~\ref{eq:rE} we obtain:
\begin{equation}
\frac{\Ml}{\Msun} = 0.123 \left(\frac{\thetaE}{\rm mas}\right)^{2}\frac{\Ds}{\rm kpc}\frac{x}{1-x} .
\end{equation}
Given the mass luminosity relations for MS stars\cite{1997MNRAS.287..402K} the upper limit for the lens distance is determined as $\Dl < 5.4$ \kpc. Using the Galactic disk models\cite{1996ApJ...467..540H}, a maximum likelihood analysis using \thetaE and \tE gives $\Dl =  5.2^{+0.2}_{-2.9}$ \kpc (90\% confidence). This implies that the lens star is an M2 - M7 dwarf with mass  $0.36^{+0.03}_{-0.28}\Msun$. The best-fitting model mass ratio, $q = 3.9\times10^{-3}$, and separation $d = 1.12$ therefore suggests that the planetary companion has mass $M_{\rm p} = 1.5^{+0.1}_{-1.2}\Mj$, at  $3.0^{+0.1}_{-1.7}$ AU from the lens star.

\section{Other events}
Other planets have been discovered by microlensing since \moaLIII. The events are summarised below.

\subsection{\ogleLXXI}
Event \ogleLXXI was a spectacular example of a high-amplification event\cite{2005ApJ...628L.109U}. The source star passed close to two caustic cusps, producing a characteristic double-peaked feature during the event maximum. The light curve was well-sampled with high accuracy throughout the event. Fitting the light curve with planetary models yielded a best-fitting model with a planetary mass ratio $q = (7.1 \pm 0.3)\times10^{-3}$ at separation $d = 1.294 \pm 0.002$ \rE. The most probable lens mass and distance was obtained from a preliminary analysis of finite source star effects during the event peak and parallax effects\cite{2002MNRAS.332..962S} seen during the event wings. The lens mass was estimated to be $0.08 \leq \Ml/\Msun \leq 0.5$ at a distance of $1.5\, \kpc \leq \Dl \leq 5\, \kpc$. These values suggest that the absolute mass of the planetary companion to be  $0.05 \leq \Mp/\Mj  \leq 4$. Assuming \Ds = 8 \kpc and \Dl = 4 \kpc, the planetary orbit radius is $1.14 \leq a \leq 2.9$ AU. These results are preliminary, pending a fuller analysis once the source amplification has returned to its baseline level.

\subsection{\ogleCCCXC}
To date, the lowest-mass planet known to orbit a main-sequence star was discovered in microlensing event \ogleCCCXC\!\cite{2006Natur.439..437B}. This event was of low maximum amplification, $\Amax \simeq 2.8$, and the planetary deviation occurred in the wings of the event.  The best-fitting light curve model is for a planet with mass ratio $q = (7.6\pm 0.7)\times10^{-5}$ at an orbit separation of $d = 1.610\pm 0.008$ \rE. The lens star mass and distance was found to be $\Ml = 0.22^{+0.21}_{-0.11}\Msun$ and $\Dl = 6.6\pm 1.0$ \kpc (68\% confidence). These values suggest that the planet has a mass of $5.5^{+5.5}_{-2.7}\Mearth$ and orbit radius $2.6^{+1.5}_{-0.6}$ AU. This remarkable event demonstrates the sensitivity of microlensing events to low mass planetary companions.

\subsection{\ogleCLXIX}
This event was identified in a timely manner as a likely high amplification event and was therefore potentially extremely sensitive to the presence of low-mass planets around the lens star. The maximum amplification for this event reached $\Amax \simeq 800$. Many observations were obtained around the peak, and a small but significant deviation from a single lens model was detected which showed the characteristics of a planetary perturbation. Subsequent analysis\cite{ogle169} showed that the light curve was best modelled assuming a planet with mass ratio $q = 8^{+2}_{-3}\times10^{-5}$ and separation $d = 1.00\pm0.02$ $(3 \sigma)$. A ML analysis yielded the lens mass and distance to be $\Ml = 0.49^{+0.23}_{-0.29}\Msun$ and $\Dl=2.7^{+1.6}_{-1.3}$ \kpc (90\% confidence). The most probable mass of the planet is therefore $\simeq 13\Mearth$, orbiting at $\simeq 2.7$ AU. This event highlights the utility of high amplification events for the detection of low-mass planets, and emphasises the need for accurate, well-sampled light curve data around the event peak. 

\subsection{\machoXCVIII}
Event \machoXCVIII was a high amplification event with $\Amax \simeq 96$ and showed a perturbation near the event peak that was consistent with a low-mass planet. The event was first analysed by Rhie\cite{2000ApJ...533..378R} \etal, and subsequently re-analysed using difference imaging photometry on an expanded dataset\cite{2002MNRAS.333...71B}. The best-fitting model from the latter analysis corresponded to a star and planet lens system. The planet mass ratio was determined to be $q = 1.3^{+0.2}_{-0.9}\times 10^{-5}$ at an orbit radius of 1.225 \rE. This corresponds to a planet with mass $(0.4 - 1.5) \Mearth$ at 2.3 AU. These values are assuming the following values of the lens and source star parameters: $\Ml = 0.3\Msun$, $\Dl = 6$ \kpc, $\Ds = 8$ \kpc and $\rho = 2\times10^{-3}$, i.e. $\Rs = \Rsun$.

The two analyses of this event strongly indicate that the planetary signal is real. Objections to this event being credited as the first detection of a planet via microlensing are largely based on the relatively low signal strength. However, some of the photometry for this event was not produced using the state-of-the-art difference imaging analysis technique\cite{2001MNRAS.327..868B}. Until these data images are analysed using the difference imaging method the results from this event will remain inconclusive.

\section{Single Lenses}
The sensitivity of high amplification events to lens system planets has been demonstrated in a number of microlensing events\cite{2001ApJ...556L.113A,2002ApJ...566..463G,2000ApJ...533..378R,2002MNRAS.333...71B,2004Sci...305.1264A,2004ApJ...616.1204Y,dong}. Given sufficient temporal coverage of the event peak, planets can be excluded from the lens star system. Planetary exclusion zones were produced for event MOA-2003-BLG-32/OGLE-2003-BLG-219 where the FWHM of the event peak was intensively monitored. The observations eliminated the presence of gas-giant and ice-giant mass planets over a wide range of orbit radii\cite{2004Sci...305.1264A,dong}. Such well-sampled high amplification light curves can help determine the fraction and character of planetary systems in the Galaxy.

\section{Discussion}

\begin{figure}[th]
\centerline{
\psfig{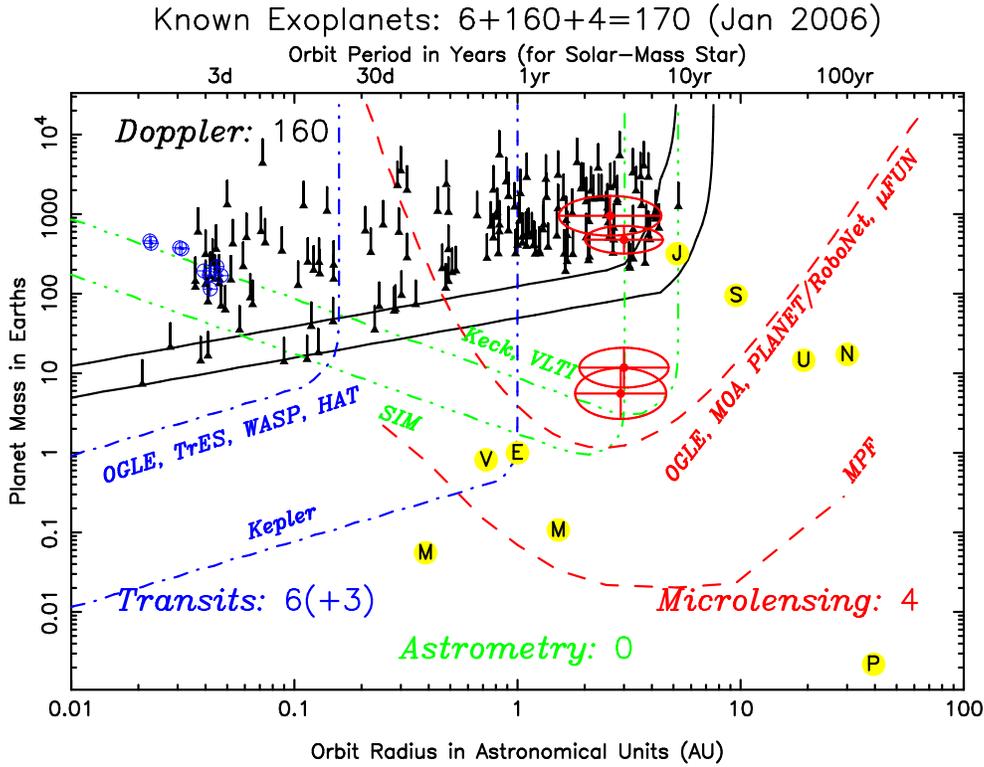}
}
\vspace*{8pt}
\caption{Extra-solar planet mass/orbit parameter space. The currently known circumstellar extra-solar planets are shown as a function of planet mass and orbit radius. Systems discovered via radial velocity measurements (black triangles), transits (blue circled plus) and microlensing (red ellipses) are shown. The theoretical detection limits of the various methods are also shown; radial velocity measurements corresponding to sensitivities 3 \mps and 1 \mps (black lines), transit detections by ground and space telescopes (blue dot-dashed), astrometry (green dot-dashed) and microlensing (red, dashed). Solar system planets are indicated with yellow dots. Figure courtesy K.~Horne.\protect\label{fig:KHogram}}
\end{figure}

\begin{figure}[th]
\psfrag{Tidal Lock Radius}[][]{\rule{30pt}{0pt}Tidal Lock Radius}
\psfrag{xlabel}[][]{Planet orbit radius (AU)\rule{0pt}{20pt}}
\psfrag{ylabel}[][]{Star mass $(\Msun)$}
\psfrag{AA}[][]{A}
\psfrag{AB}[][]{B}
\psfrag{AC}[][]{C}

\centerline{
\psfig{file=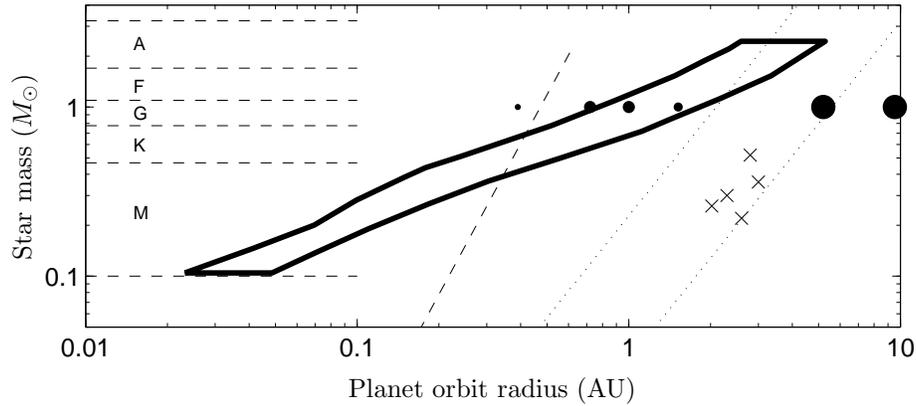,width=1.0\textwidth}
}
\vspace*{8pt}
\caption{Microlensing planet environments. The most likely orbit radius and host star mass for the planets listed in Table~\ref{tab:planets} are indicated with cross symbols. The habitable zone (thick lines) and tidal lock radius (oblique dashed line) determined by Kasting\protect\cite{1993Icar..101..108K} \etal is shown, along with the inner six Solar system planets. Spectral classifications\protect\cite{Allen} are indicated by horizontal dashed lines. The lensing zone ($0.6 \lesssim d \lesssim 1.6$) (dotted lines) of Park\protect\cite{Park} \etal is also shown. Note that the lensing zone for a space telescope will be wider, see Sec.~\ref{sec:MPF}.
  \protect\label{fig:habzone}}
\end{figure}

The parameters for the three planetary microlensing events are summarised in Table~\ref{tab:planets}. The mass and orbit radius of all known extra-solar planets are shown in Fig.~\ref{fig:KHogram},  along with the theoretical limits of each of the detection methods used. Microlensing is currently detecting planets in a previously unreachable region of the planetary mass-radius space. Microlensing is returning detections of planets with masses approaching that of Earth. Fig.~\ref{fig:habzone} shows the mass-orbit plane for the set of microlensing planets listed in Table~\ref{tab:planets} with the solar system planets for comparision, along with the theoretical habitable zone, following Kasting\cite{1993Icar..101..108K} \etal Since most of the lens system planets are M-dwarf stars, the planets discovered by microlensing are well removed from the habitable zone for their host stars. The lensing zone, the range of planet-star separation where planets are most likely to produce detectable signals, only overlaps with the habitable zone for lens masses $\Ml>\Msun$, see Fig.~\ref{fig:habzone}. A space telescope will have a higher sensitivity to habitable planets via microlensing, see Sec.~\ref{sec:MPF}.

\begin{table}
\tbl{\label{tab:planets}Planetary microlensing event parameters}
{\begin{tabular}{cccccccccc} \toprule \toprule
\
&
&
&
\multicolumn{2}{c}{Source Star}&
\multicolumn{3}{c}{Lens Star$^{\rm a}$}&
\multicolumn{2}{c}{Planet}\\
Event &
\Amax &
\tE (d)&
Sp$^{\rm b}$ &
\rhoStar &
Sp$^{\rm b}$ &
\Ml (\Msun)&
\Dl (\kpc)&
$q$ $(\times10^{-3})$ &
$d$ $(\rE)$
\\ \toprule
O235/M53&
7.5&
61.5&
G&
0.00096&
M2-M7 &
$0.36^{+0.03}_{-0.28}$&
$5.2^{+0.2}_{-2.9}$&
$3.9^{+1.1}_{-0.7}$ &
1.120(7)  \tabStrut \\
O71$^{\rm c}$ &
42 &
70.9 &
? &
? &
? &
0.08 -- 0.5&
1.5 --  5&
$7.1(3)$ &
$1.294(2)$  \tabStrut \\
O390 &
2.8&
11.03&
G4$^{\rm d}$ &
0.026 &
M &
$0.22^{+0.21}_{-0.11}$ &
$6.6^{+1.0}_{-1.0}$ &
0.076(7) &
1.610(8)  \tabStrut \\
O169 &
$\simeq 800$ &
$44$ &
? &
$\simeq 0.0005$ &
M &
$0.49^{+0.23}_{-0.29}$ &
$2.7^{+1.6}_{-1.3}$ &
$0.08^{+0.02}_{-0.03}$ &
$1.00(2)$ \tabStrut \\
98-35 &
96 &
27.7 &
? &
0.002$^{\rm e}$ &
? &
0.3$^{\rm e}$ &
6.0$^{\rm e}$ &
$0.013^{+0.002}_{-0.009}$ &
$1.225^{\rm f}$  \tabStrut \\

\botrule
\end{tabular}}
$^{\rm a}$Quoted confidence limits are 90\% for events \moaLIII and \ogleCLXIX, and 68\% for event \ogleCCCXC. $^{\rm b}$Luminosity class V unless specified otherwise.  $^{\rm c}$Preliminary analysis. $^{\rm d}$Luminosity class III. $^{\rm e}$Assumed value. $^{\rm f}$Degenerate solution at $1/d$ = 0.81 \rE.

\end{table}

\section{Future prospects}
Recent progress has been made in the establishment or improvement of several microlensing experiments. We discuss some of the advances made by the microlensing community toward detecting low-mass extra-solar planets.

\subsection{MOA-2}

The MOA collaboration has recently commissioned a new 1.8 m telescope dedicated to detecting microlensing events in the Galactic bulge and the Magellanic clouds\cite{MOA2pre}. The telescope optics gives a field of view of 2.2 square degrees. The telescope is currently undergoing final calibration tests and is expected to be fully operational in survey mode by March 2006. The detection of extra-solar planets is most efficient toward the Galactic bulge fields. The MOA-II telescope will be used to survey 23 fields toward the bulge several times per night. The increased photometric accuracy produced by the large primary of the MOA-II telescope and the increase in field sampling rate will improve the detection efficiency of high-amplification events and low-mass planets. 

\subsection{Earth-Hunter network}

A new method for maximising the discovery rate of low-mass planets through microlensing has recently been proposed\footnote{A. Gould, 5-6th November 2005, Max Planck Institute for Astronomy, Heidelberg.}. Four 2 m class telescopes spaced longitudinally around the Earth, each with a field-of-view of four square degrees, would observe only a few (2, 3 or 4) fields per night.  The proposed network will detect around 7000 events per year. Most of these events will be of low maximum amplification. Observing only a few fields per night would produce light curves with very fine time sampling and the use of 2 m class telescopes in locations with good seeing would result in high quality light curves. The increased event rate combined with the high time resolution and improved photometry will result in a higher detection rate of low-mass planets\cite{Nagoya}.

\subsection{Microlensing Planet Finder}
\label{sec:MPF}
The Microlensing Planet Finder (MPF) is a proposed space telescope dedicated to microlensing observations\cite{2004SPIE.5487.1453B,2003SPIE.4854..141Bs,2002ApJ...574..985B}. The MPF would have the obvious advantage of being able to observe an event continuously, uninterrupted by weather or daylight. Improved photometric accuracy is another strong advantage and would allow the detection of planets with masses less than that of Mercury. In a large sample of microlensing events, there will be some where the lens and source are quite close to each other. The physical scales probed in such events will therefore be smaller. The high sensitivity of the MPF telescope combined with its ability to observe a large number of events means that the MPF would be capable of detecting Earth-mass planets in the habitable zone of G and K dwarf stars.

\section{Conclusion}

Microlensing was suggested as a method for detecting extra-solar planets over a decade ago. Four planets have been detected via microlensing to date, including the lightest planet found orbiting a main-sequence star.  Microlensing is a uniquely sensitive method for the discovery of low-mass extra-solar planets. The four planets discovered to date through microlensing have masses ranging from $5.5\Mearth$ to $3\Mj$ and orbit their host star at distances of $\sim$ few AU. Microlensing is allowing a previously undetectable class of planets to be discovered.

Improvements in survey and follow-up instrumentation and operation will increase the discovery rate of low-mass planets, leading to estimates of the Galactic planetary mass-function. Many more planets will be discovered via microlensing. New ground and space telescopes will have higher sensitivity to low-mass planets than current instrumentation. In particular, a space telescope such as the proposed MPF mission will be sensitive to habitable Earth-like planets around Sun-like stars. Within the next ten years we can expect that dozens of extra-solar planets will have been discovered via microlensing, possibly some very similar to Earth.

\section{Acknowledgements}
Thanks to Shude Mao, Phil Yock and Dave Bennett for many helpful discussions and for their comments on this manuscript. NJR is supported by a PPARC postdoctoral grant. This work was partially supported by the European Community's Sixth Framework Marie Curie Research Training Network Programme, Contract No. MRTN-CT-2004-505183 `ANGLES'.

\bibliographystyle{nix3}

\bibliography{microlensing}

\end{document}